\documentclass[pdflatex,sn-mathphys-num]{sn-jnl}% Math and Physical Sciences Numbered Reference Style
\usepackage{graphicx}%
\usepackage{multirow}%
\usepackage{amsmath,amssymb,amsfonts}%
\usepackage{amsthm}%
\usepackage{mathrsfs}%
\usepackage[title]{appendix}%
\usepackage{xcolor}%
\usepackage{textcomp}%
\usepackage{manyfoot}%
\usepackage{booktabs}%
\usepackage{algorithm}%
\usepackage{algorithmicx}%
\usepackage{algpseudocode}%
\usepackage{listings}%
\usepackage{physics}

\theoremstyle{thmstyleone}%
\theoremstyle{thmstyletwo}%

\theoremstyle{thmstylethree}%

\begin{document}

\title[Rotating Quantum Droplets in Low Dimensions]{Rotating Quantum Droplets in Low Dimensions}


\author*[1]{\fnm{K.} \sur{Hernández}}\email{kevinhernandezbel@hotmail.com}

\author[2]{\fnm{E.} \sur{Castellanos}}\email{elias.castellanos@tec.mx}

\affil*[1]{\orgdiv{Escuela de física, Facultad de Ciencias Naturales y Matemáticas}, \orgname{Universidad de El Salvador}, \orgaddress{\street{Final de Av. Mártires y Héros del 30 de julio}, \city{San Salvador}, \postcode{111009}, \state{San Salvador}, \country{El Salvador}}}

\affil[2]{\orgdiv{Escuela de Ingeniería y Ciencias}, \orgname{Tecnológico de Monterrey}, \orgaddress{\street{Carr. Lago de Guadalupe Km. 3.5}, \city{Estado de México}, \postcode{52926}, \country{México}}}


\abstract{Quantum droplets formed by rubidium, lithium, and sodium atoms have been analyzed in this paper by using a logarithmic-type Gross-Pitaevskii equation. Variational methods and numerical techniques were employed to solve the corresponding nonlinear equations. A disk-shaped Bose-Einstein condensate was analyzed to assess its radial evolution. Additionally, free expansion under rotation of the BEC was studied. Compression and expansion around the equilibrium radius were observed in different scenarios, predicting self-confinement, which implies the formation of quantum droplets originating from a BEC state. Briefly, the physical aspects of the system and the possible formation of Bose-nova effects are discussed.}




\maketitle

\section*{Introduction}
 
The emergence of a new phase transition from a Bose-Einstein condensate (BEC) to a quantum state with liquid-like characteristics has become a rapidly growing area of research in recent years. The BEC was first achieved in 1995 using magneto-optical traps, as in the cases of ${}^{87}\mathrm{Rb}$ and ${}^{23}\mathrm{Na}$ gases \cite{RevModPhys.71.463}, where atoms were confined in magnetic traps and cooled to extremely low temperatures, on the order of fractions of a microkelvin. Numerous articles discuss the Bose-Einstein condensate both theoretically and experimentally. BECs have been analyzed in lower dimensions due to the peculiarity that a homogeneous BEC cannot exist in one or two dimensions; however, it was discovered that by reducing the number of atoms, it is possible to create a 2D BEC using magnetic traps \cite{gorlitz2001realization}. BEC formation has also been studied in molecular systems \cite{wynar2000molecules}. Furthermore, different groups have used optical dipole traps to obtain BECs, eliminating the need for magnetic confinement and enabling studies of BEC interactions within optical lattices \cite{PhysRevLett.80.2027, RevModPhys.78.179, PhysRevLett.79.553}.

Quantum droplets in Bose-Einstein condensates are small self-bound clusters of atoms stabilized by the balance between attractive and repulsive forces \cite{cabrera2018quantum}, also a finite volume of liquid is what different authors called quantum droplets in Ref. \cite{zloshchastiev2012volume}, while analyzing the superfluidity of liquid helium, which serves as the prime example of a quantum fluid, where the formation of droplets relies on the balance between attractive forces, which bind the atoms together, and repulsive forces, which prevent collapse. In helium droplets, the repulsion is primarily governed by the interatomic potential \cite{Zloshchastievdilute, zloshchastiev2012volume, vowe2020detecting}, therefore, Ref. \cite{Zloshchastievql} uses “transcendental condensates” for stable quantum liquids from BEC. Ref. \cite{viefers2008quantum} provides a review of both theoretical and experimental advances in quantum Hall physics in cold bosonic gases. One of the starting points of our work is the study of BEC dynamics in a rotating frame, as we aim to introduce the concept of rotation in quantum droplets. In this context, certain dynamical instabilities have been observed \cite{RevModPhys.81.647, stock2005bose, sinha2001dynamic, recati2001overcritical}. Furthermore, as the number of vortices increases, the Bose-Einstein condensate undergoes a transition to a quantum Hall-like state \cite{PhysRevLett.87.060403, weiler2008spontaneous, svidzinsky2000dynamics}.

Another fundamental aspect of this work involves the study of nonlinear Schrödinger equations with a particular focus on the logarithmic Schrödinger equation (LogSE) as analyzed in Ref. \cite{BIALYNICKIBIRULA197662}. The logarithmic Schrödinger equation (LogSE) has found applications in several areas of physics, including quantum mechanics, optics, nuclear physics, superfluidity, and notably in the study of Bose-Einstein condensates \cite{d2014logarithmic, scott2018solution, bouharia2015stability, liverts2008approximate, hefter1985application, vowe2020detecting, avdeenkov2011quantum}.

In a dipolar BEC without quantum fluctuations, if the net interaction (the sum of the contact and dipolar interactions) is predominantly attractive, the system would collapse due to the strong interparticle attraction. However, recent experiments have shown that the formation of self-stabilizing quantum droplets remains stable, despite strong attractive forces. This stability is attributed to an additional term arising from quantum fluctuations \cite{PhysRevLett.116.215301, kadau2016observing, PhysRevLett.95.200404}. Quantum fluctuations in dipolar quantum droplets emerge from the Heisenberg uncertainty principle and quantum correlations beyond the mean-field approximation.

A stable quantum droplet was reported in Ref. \cite{PhysRevLett.116.215301}, where stability was achieved via quantum fluctuations using dipolar atoms such as ${}^{164}$Dy. In Ref. \cite{kadau2016observing}, ferrofluids were previously studied, and an unstable quantum droplet was observed. Initially, these binary systems of two interacting BECs were often treated as bright solitons. Other works have similarly interpreted the combination of two BECs as solitonic structures, although they are not solitons in the strict sense \cite{PhysRevLett.95.200404}.

The Lee-Huang-Yang (LHY) correction was also applied to describe a hard-sphere behavior in BECs \cite{PhysRev.106.1135, dong2020multi}. Furthermore, alternative types of quantum fluctuations, such as those arising from three-body interactions defined through energy density or energy functionals, have been explored, and different theoretical approaches can be reviewed in Ref. \cite{RevModPhys.75.121}.
When a BEC is at rest, all atoms occupy the same lowest energy quantum state, showing coherent behavior at the macroscopic level. However, by imposing a rotation on it, vortices are generated, i.e., regions in which the density of atoms is reduced to zero and in which the phase of the quantum wave function varies in a circular pattern \cite{RevModPhys.81.647, bao2006efficient, bao2007ground}. In a previous work, the stability conditions of a one-dimensional quantum droplet were analyzed using the GPE with a logarithmic potential \cite{rodriguez2021oscillating}. 

The purpose of this work is to analyze the dynamics of a quantum droplet using the Gross-Pitaevski equation with a logarithmic potential in low dimensions within a rotation framework.
This work is organized into several sections:
Section 1 describes the different states of the literature regarding Bose-Einstein condensates under rotational dynamics, analyzing several physical and mathematical theories.
Section 2 introduces the reader to quantum droplets, particularly on their modeling through the Gross-Pitaevskii equation. Stability and confinement criteria are discussed. In this section, a full three-dimensional numerical development is carried out to evaluate the expansion dynamics. However, a disk-shaped configuration is also considered to define free parameters and control their values in order to identify stability regions.
The paper concludes with the implementation of rotation in the system. The numerical calculations presented can be generalized to other systems, although $^{87}Rb$ is mainly used here to analyze atomic stability.

\section{Bose-Einstein condensates in Rotation}
A rotating Bose-Einstein condensate (BEC) is a quantum state of matter that exhibits unique properties when external angular momentum is applied. Rotation in a BEC induces behaviors and patterns in the distribution of its atoms, affecting its structure and dynamics \cite{RevModPhys.81.647}. A BEC is described as the macroscopic wave function $\psi\qty(r)=\sqrt{\rho \qty(r)}e^{i\phi\qty{r}}$ obtained as a solution of the GPE, where $\rho$ and $\phi$ are the spatial density and phase of the fluid. However, there are restrictions on the spatial density of the gas, since in areas where $v=\frac{\hbar}{M} \grad \phi $, with M the mass of a particle and $\hbar$ Planck's constant, it implies that $\curl v=0$. In other words, the circulation of the velocity field is quantized through vortices \cite{stock2005bose}. In addition, $\rho$ could vanish in singular points or lines also  \cite{RevModPhys.81.647}.

In Ref. \cite{PhysRevLett.84.806}, the authors have reported the formation of vortices in a gaseous Bose-Einstein condensate when it is
stirred by a laser beam, which produces a slight rotating anisotropy. Therefore, it implies that
\begin{equation}
i\hbar \frac{\partial \Psi(\mathbf{r}, t)}{\partial t} = \left[ H - \Omega (t)L_z \right] \Psi(\mathbf{r}, t)
\label{Eq:BECROT}
\end{equation}

\noindent where $H= -\frac{\hbar^2}{2m} \nabla^2 + V_{\text{trap}}(\mathbf{r,t}) + g|\Psi(\mathbf{r}, t)|^2$ is the Hamiltonian in the absence of rotation.  $L_z$ is the total orbital angular momentum along the rotation axis. 
 The term \( i \hbar \frac{\partial \Psi(\mathbf{r}, t)}{\partial t} \) indicates the time evolution of the condensate's wave function while  \( -\frac{\hbar^2}{2m} \nabla^2 \) is the kinetic energy term with \( \nabla^2 \) the Laplace operator. The term \( V_{\text{trap}}(\mathbf{r}) \) represents the external trapping potential that confines the condensate defined as $V_{trap} =\frac{m\omega^2}{2}  \qty(x^2+y^2+z^2) $. Additionally, $ g |\Psi(\mathbf{r}, t)|^2$ is the nonlinear interaction, where \( g \) is the interaction strength, and \( |\Psi|^2 \) represents the condensate density. Finally, \( -\Omega L_z \),  is the rotational contribution, where \( \Omega \) is the angular velocity and \( L_z \) is the angular momentum operator along the \textit{z}-axis:
    \[
    L_z = -i\hbar \left( x \frac{\partial}{\partial y} - y \frac{\partial}{\partial x} \right).
    \]
To obtain the main features associated with a BEC in rotation, we can see in Ref. \cite{PhysRevLett.87.190402}, the GPE for BEC in rotation as shown in Eq. \ref{Eq:BECROT}. The time-dependence of $\Omega(t)$ vanishes when time-dependent frequencies such as pulses or other electromagnetic phenomena are not used, therefore, $\Omega(t)=\Omega$ and $V_{trap}(r,t)=V_{trap}(r)$, or using Ref. \cite{RevModPhys.81.647}, as a frame in rotation. 

\section{Quantum Droplets from Log-GPE model}
A quantum droplet is a self-bound Bose-Einstein condensate (BEC) stabilized by internal interactions. In this work, we show that such droplets can emerge from a simple BEC by introducing a logarithmic potential that models many-body effects. Ref. \cite{scott2019resolving, Zloshchastievdilute} one can find one of the pioneers in finding a cubic dependence of the Gross-Pitaevski equation using the experimental behavior of the speed of sound in liquid helium and dilute-BEC. To achieve droplet formation, an anisotropic trapping potential is preferred, enabling effective low-dimensional interactions with the logarithmic term. Alternatively, turning off the trap along the axial direction allows the many-body interactions to provide self-confinement, signaling the emergence of quantum droplets, furthermore, the time-free expansion analysis of a BEC using GPE was studied in Ref. \cite{vowe2020detecting}, using variational methods and a strong time dependence on the BEC radius, because it is a symmetric expansion, they also found oscillations under appropriate parameters. Thus, it is unnecessary to rely on dipolar BECs or BEC mixtures.

 In a previous study \cite{rodriguez2021oscillating}, the stability of a one-dimensional quantum droplet was analyzed using the Gross-Pitaevskii equation (GPE) with an additional logarithmic term. The GPE used is given in Eq. \ref{Eq:QD}:
\begin{equation}
i\hbar \frac{\partial \Psi(\mathbf{r}, t)}{\partial t} = \left[ -\frac{\hbar^2}{2m} \nabla^2 + V_{\text{trap}}(\mathbf{r,t}) + g_{3D}|\Psi(\mathbf{r}, t)|^2 - \beta \ln\qty[\alpha ^3 |\Psi(\mathbf{r}, t)|^2] \right] \Psi(\mathbf{r}, t),
\label{Eq:QD}
\end{equation}    

\noindent where $g_{3D}=\frac{4\pi \hbar^2}{m}a_s$ is the interaction strength between any pair of bosons in three-dimensions, with $a_s$ the s-wave scattering length of the corresponding gas while $\beta$ and $\alpha^3$  measure the strength of the nonlinear logarithmic interaction.  

At this point, it is important to remark some relevant characteristics associated with the system, that is:
\begin{itemize}
    \item Self-confinament: The quantum droplets under certain conditions can be stabilized; this depends strongly on the number of bosons as Ref. \cite{donley2001dynamics}.
    \item Under free expansion, that is, when the magnetic trap is turned off in the \textit{z} direction, it can expand or oscillate, depending on the number of particles, in addition, there is no dependence on $\alpha$ as Ref. \cite{rodriguez2021oscillating, Zloshchastievdilute}.
    \item  The $\beta$ parameter only acts as an energy balance that allows quantum droplet formation on its own. Quantum droplet formation by LHY or Binary BEC has been reported as \cite{cabrera2018quantum, khan2022quantum, Zloshchastievdilute, vowe2020detecting}, but quantum droplet formation by self-confinement also exists with a non zero $\beta$ as was reported in \cite{rodriguez2021oscillating, vowe2020detecting}.
\end{itemize}

\subsection{BEC using LogSE in spherical coordinates in non-rotating frame \label{sec: BEC3D-NotRot}}
The energy functional for a BEC in LogSE context \cite{zloshchastiev2012volume, rodriguez2021oscillating} is given by:
 \begin{eqnarray}
 \label{eq:E}
E[\Psi] =\int d\mathbf{r}^3 \left( \frac{\hbar^2}{2m} |\nabla \Psi|^2 + V_{\text{trap}} |\Psi|^2 + \frac{g_{3D}}{2} |\Psi|^4 - \beta |\Psi|^2\ln\left(\alpha^3 |\Psi|^2\right) \right).    
 \end{eqnarray}

\noindent If the condensate wave function depends only on the radial component, it naturally implies that the derivatives of the other components give zero as a result. To calculate the total energy in Eq.~\ref{eq:E}, we use the ansatz $\Psi(r, t) = A e^{-r^2 / 2a^2} e^{-iEt/\hbar}$, where $|\Psi|^2 = |\psi(r)|^2$, and impose the normalization condition $\int |\Psi|^2\, d^3r = N.
$. This yields the normalization constant: $A = \sqrt{\frac{N}{\pi^{3/2} a^3}}$. Spherical symmetry will be used, where $r$ represents the distance from the center of the system to some point P, $\theta$ represents the angle around \textit{z}-axis while $\phi$ represents the angle between the radial line and the \textit{z}-axis, so, the small volume is expressed as $dv=r^2 \sin \theta dr d\theta d\phi$, and the log term is reduced as
\begin{equation}
    -\beta\ln (\alpha^3\abs{\psi(r)}^2)=-\beta \qty[\ln(\frac{\alpha^3 N}{\pi^{3/2} a^3})-\frac{r^2}{a^2}].
\end{equation}
    
\noindent In order to obtain the total energy of the system, we analyze the following results:

 \begin{equation}
E_{kin} = \int d\mathbf{r}^3 \left( \frac{\hbar^2}{2m} |\nabla \Psi|^2 \right)=\frac{3}{4}\frac{\hbar^2 N}{ma^2}
\end{equation}

\noindent and subject to a harmonic trap $V_{trap}=\frac{1}{2}m\omega^2 r^2$:
 \begin{equation}
E_{trap} = \int d\mathbf{r}^3 \left(V_{\text{trap}} |\Psi|^2  \right)=\frac{3}{4}N m \omega^2 a^2.
\end{equation}
\noindent The interaction term is easy to calculate:
\begin{equation}
E_{int}= \int d\mathbf{r}^3 \left(\frac{g_{3D}}{2} |\Psi|^4 \right)=\frac{g_{3D}N^2}{4\sqrt{2 \pi^3} a^3},
\end{equation}
\noindent and finally the contribution of the log term is reduced to

 \begin{equation}
E_{log} = \int d\mathbf{r}^3 \left( - \beta |\Psi|^2\ln\left(\alpha^3 |\Psi|^2\right) \right)=-\frac{\beta N}{2}\qty[2\ln(\frac{\alpha^3 N}{\pi^{3/2} a^3})-3].
\end{equation}

\noindent Thus, the total energy becomes:
\begin{equation}
    E=\frac{3}{4}\frac{\hbar^2 N}{ma^2}+\frac{3}{4}N m \omega^2 a^2+\frac{g_{3D}N^2}{4\sqrt{2 \pi^3} a^3}-\frac{\beta N}{2}\qty[2\ln(\frac{\alpha^3 N}{\pi^{3/2} a^3})-3].
    \label{Eq:EnergyBEC-Spherical}
\end{equation}

\noindent If we were dealing with a 3D quantum harmonic oscillator, only the first two terms on the right-hand side of Eq.~\ref{Eq:EnergyBEC-Spherical} would contribute to the energy minimization via the variational principle. In that case, the optimal parameter is $a = \sqrt{\frac{\hbar}{m\omega}}$, and the minimum energy is $E_{\text{min}} = \frac{3}{2} \hbar \omega$. In Ref.~\cite{PhysRevA.87.053614}, the scattering length \(a_s\) is calculated for \({}^{87}\mathrm{Rb}\) to determine the interaction strength \(g_{3D}\), yielding a value of \(98\,r_0\), where \(r_0\) is the Bohr radius. In Ref.~\cite{rodriguez2021oscillating}, values for \({}^{27}\mathrm{Na}\) and \({}^{7}\mathrm{Li}\) are reported as \(53.65\,r_0\) and \(-27.3\,r_0\), respectively. For practical purposes in our plots, the parameters \(\beta\) and \(\alpha\) will be kept fixed. Additionally, we introduce the dimensionless variable \(r = a/a_0\), where \(a_0\) denotes the size of the BEC in the absence

\begin{figure}[htbp!]
\centering
   \includegraphics[width=0.8\linewidth]{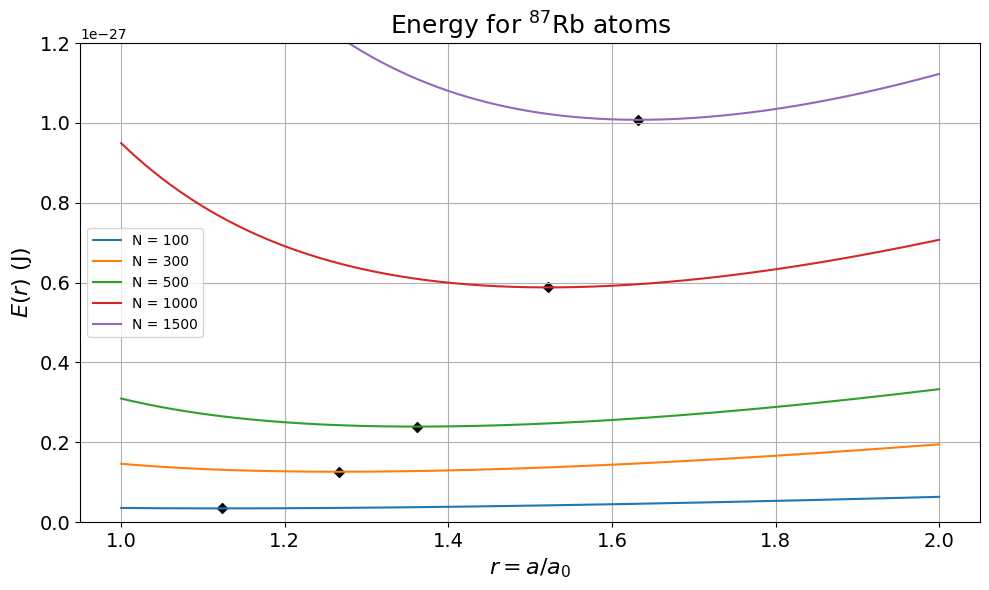}
    \caption{Plots of Eq. \ref{Eq:EnergyBEC-Spherical} for $^{87}Rb$. The black dots on the curve indicate the radii corresponding to the minimum energy. It is observed that the radius increases as the number of atoms increases.}
    \label{fig:QD-RB-SC-Energy}
\end{figure}

\begin{figure}[htbp!]
\centering
\includegraphics[width=0.8\linewidth]{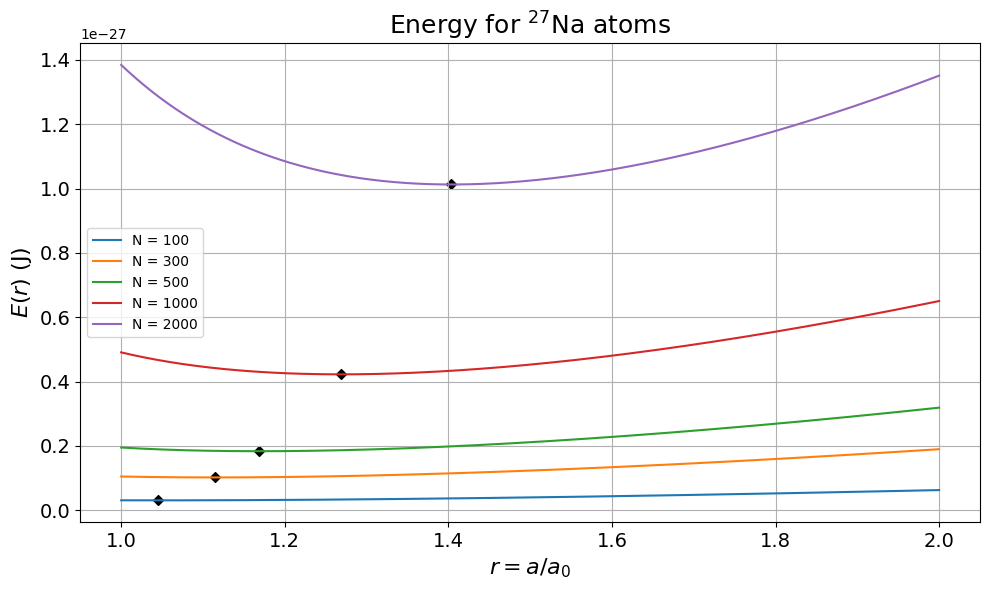}
\caption{Plots of Eq. \ref{Eq:EnergyBEC-Spherical} for $^{27}Rb$. The black dots on the curve indicate the radii corresponding to the minimum energy. It is observed that the radius increases as the number of atoms increases i.e., the same behavior as the $^{87}Rb$ atom.}
\label{fig:QD-NA-SC-Energy}
\end{figure}

\begin{figure}[htbp!]
\centering
\includegraphics[width=0.8\linewidth]{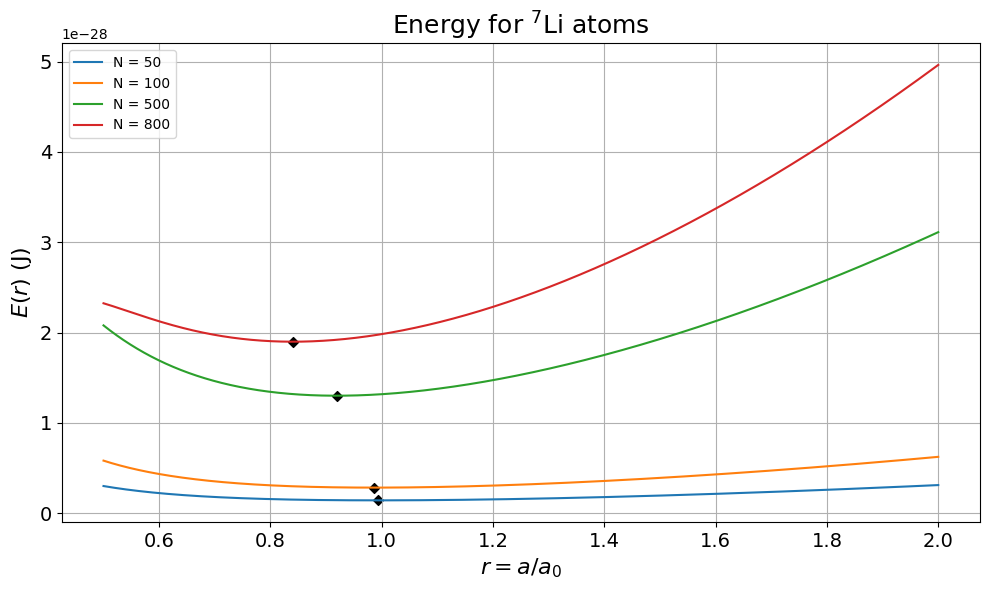}
\caption{Plots of Eq. \ref{Eq:EnergyBEC-Spherical} for $^{7}Li$. The black dots on the curve indicate the radii corresponding to the minimum energy. It is observed that the radius decreases as the number of atoms increases, this is because the scattering length of $^{7}Li$ atoms is negative unlike $^{27}Na$ and $^{87}Rb$.}
\label{fig:QD-LI-SC-Energy}
\end{figure}
The plots in Figs.~\ref{fig:QD-RB-SC-Energy}, \ref{fig:QD-NA-SC-Energy}, and \ref{fig:QD-LI-SC-Energy} were generated using Eq.~\ref{Eq:EnergyBEC-Spherical}, with the parameters \(\beta = 1 \times 10^{-3} \hbar \omega\), \(\alpha = 1\), and \(\omega / 2\pi = 300\,\text{Hz}\), considering a radius range from 1 to 2 times \(a_0\). The black dots indicate the radius that minimizes the energy, corresponding to the equilibrium configuration.

Fig. ~\ref{fig:QD-RB-SC-Energy} shows the result for \({}^{87}\mathrm{Rb}\), a relatively heavy atom with a scattering length \(a_s = 98\,r_0\), while Fig.~\ref{fig:QD-NA-SC-Energy} presents the case for \({}^{27}\mathrm{Na}\), using \(a_s = 53.65\,r_0\). In both cases, the equilibrium radius is larger than the BEC radius without the logarithmic interaction. In contrast, Fig.~\ref{fig:QD-LI-SC-Energy} shows the case for \({}^{7}\mathrm{Li}\), where the equilibrium radius is smaller than the non-interacting radius. This is due to the negative scattering length, indicating that the corresponding interaction term is purely attractive.

An additional point of interest is the estimation of the average radius of the BEC under these conditions. The quantity \(\ev{r^2}{\Psi}\) serves as the appropriate measure, which can be computed using the proposed ansatz and assuming \(a \approx a_0\) as a first-order approximation.

Thus, we obtain:

\begin{eqnarray}
\label{eq:2}
    \ev{r^2}{\Psi}=\int \abs{\Psi}^2 r^2 dr^3=\frac{3}{2}N a_0^2=\frac{3}{2}\frac{N\hbar}{m\omega},
\end{eqnarray}
and represents the mean square radius of the BEC, which reflects the average spatial spread squared. We noticed that Eq.\,\ref{eq:2} scales as \( N \), the number of particles, as well with the trap frequency \( \omega \), where a lower frequency indicates a wider spread due to weaker confinement as
\begin{eqnarray}
\label{eq:3}
\ev{r}{\Psi}=\int \abs{\Psi}^2 r dr^3=\frac{2}{\sqrt{\pi}}N a_0
\end{eqnarray}
\noindent Eq.\,\ref{eq:3} represents the average radial distance of particles in the BEC from the trap center. This quantity scales with both \( N \) and \( a_0 \), indicating that the size of the condensate increases as the number of particles \( N \) grows or as the trap frequency \( \omega \) decreases since a lower \( \omega \) leads to a larger \( a_0 \)~\cite{gorlitz2001realization, RevModPhys.81.647, PhysRevLett.116.215301}.

\subsection{Free expansion of the BEC 3D in LogSE context}
In Ref.~\cite{pethick2008bose}, the theory of free expansion of a Bose-Einstein condensate (BEC) is developed, and in Ref.~\cite{rodriguez2021oscillating}, this framework is applied to demonstrate the existence of quantum droplets. To analyze time-dependent potentials or study the time evolution of the BEC, it is necessary to reformulate the ansatz \(\Psi(r, t) = A e^{-r^2/2a^2} e^{-iEt/\hbar}\) in the more general form \(\Psi(r, t) = f(r, t) e^{i\eta(r)}\).

In this hydrodynamic representation, the equations take the form \(N = |f|^2\) and \(\mathbf{v} = \frac{\hbar}{m} \nabla \eta\), where \(N\) is the BEC density, \(m\) is the atomic mass, and \(\eta\) is the phase of the wavefunction. Under this formulation, an additional kinetic term associated with the fluid velocity appears in the total energy expression.

\begin{equation}
    E_{total}=E_{flow}+E_{kin}+E_{trap}+E_{int}+E_{log},
\end{equation}
only the first terms differs of Eq. \ref{Eq:EnergyBEC-Spherical} and its definition is:
\begin{equation}
    E_{flow}=\frac{\hbar^2}{2m}\int \abs{\psi}^2 \abs{\grad \eta}^2 dr^3.
\end{equation}

\noindent Under free expansion the radial speed can be modelate as $v(r)=r\frac{\dot{a}}{a}$, then is straightforward to obtain $E_{flow}$, with the result:
\begin{equation}
    E_{flow}=\frac{\hbar^2}{2m}\int \abs{\psi}^2 \abs{\grad \eta}^2 dr^3=\frac{3}{4}mN\dot{a}^2.
\end{equation}
\noindent For practical purposes, we start with the equilibrium radius in the absence of logarithmic interactions, i \( a = \sqrt{\frac{\hbar}{m\omega}} \). When the harmonic trap is turned off, the system has an initial energy \( E_0 \) at time \( t = 0 \). At later times \( t \), the energy becomes \( E \), which corresponds to a modification of the equilibrium radius. In other words, we assume energy conservation. Consequently, we obtain

%From here, we will go by conservation of energy from when the trap is turned off until a time t:

\begin{eqnarray}
    \frac{3}{4}mN\dot{a}^2+\frac{3}{4}\frac{\hbar^2 N}{ma^2}&+&\frac{g_{3D}N^2}{4\sqrt{2 \pi^3} a^3}-\frac{\beta N}{2}\qty[2\ln(\frac{\alpha^3 N}{\pi^{3/2} a^3})-3]= \nonumber \\
    &&\frac{3}{4}\frac{\hbar^2 N}{ma_0^2}+\frac{g_{3D}N^2}{4\sqrt{2 \pi^3} a_0^3}-\frac{\beta N}{2}\qty[2\ln(\frac{\alpha^3 N}{\pi^{3/2} a_0^3})-3].
\end{eqnarray}

\noindent Thus, for $\dot{a}^2$, we get:
\begin{eqnarray}
    \dot{a}^2=\frac{\hbar^2}{m^2}\qty(\frac{1}{a_0^2}-\frac{1}{a^2})+\frac{g_{3D}N}{3m\sqrt{2\pi^3}}\qty(\frac{1}{a_0^3}-\frac{1}{a^3})+\frac{4\beta}{m}\ln \qty(\frac{a_0}{a}).
    \label{Eq:dota-3DQD}
\end{eqnarray}
The Eq. \ref{Eq:dota-3DQD} represents the square of the velocity of the free expansion. The idea is to find a region where the velocity increases and then gradually decreases. This region would imply that the velocity (and also the size of the system) oscillates around the equilibrium point. Therefore, if the velocity oscillates, there is a possibility that there is a region that depends on parameters such as the mass, the interaction parameters, that is the s-wave scattering length, and $\beta$. There can be 4 solutions for the BEC radius: a) a constant radius, b) an expanding radius or c) an oscillation around an equilibrium radius, d) an implosion, i.e. a radius that can contract from its equilibrium point. Notice that when $g_{3D}=0$ and $\beta=0$, we recovered the usual equation of motion related to the expansion that corresponds to the velocity predicted by the Heisenberg uncertainty principle:

\begin{eqnarray}
    a^2=a_0^2+\qty(v_0t)^2, v_0=\frac{\hbar}{ma_0}.
    \label{eq:adot-Spherical}
\end{eqnarray}
Eq. \ref{eq:adot-Spherical} suggests that the free expansion of the BEC would continue indefinitely. However, the interaction parameters play a crucial role: depending on their values, the system may stop expanding, contract, or exhibit an oscillatory behavior.

\begin{figure}[htbp]
    \centering
    \includegraphics[width=0.8\linewidth]{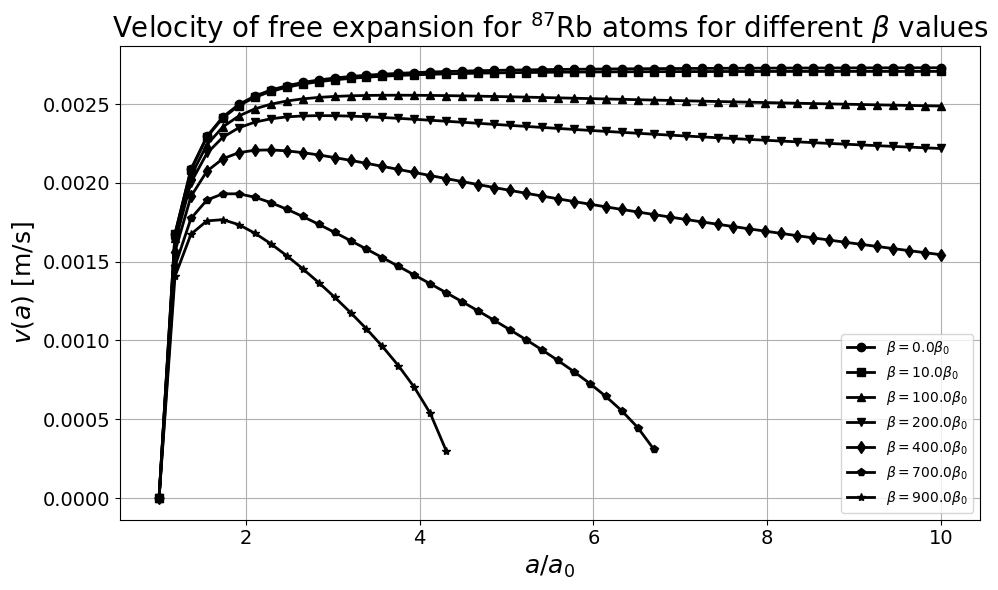}
    \caption{Plots of the Eq. \ref{Eq:dota-3DQD} using $^{87}Rb$ properties as: $a_s=98r_0$, $m = 1.44316060\times 10^{-25} kg$, $\omega/ 2 \pi =300 Hz$, $\beta=1\times 10^{-3} \hbar \omega$,  $a_0 = \sqrt{\hbar / (m\omega)}$, $N=1000$ atoms with several values of $\beta$. The velocity of BEC expansion varies according to the size of the BEC, notably when the logarithmic interaction is larger, it tends to reach a maximum point and then decreases for the other cases it expands abruptly and then expands steadily.}, 
    \label{fig:Va_RB}
\end{figure}

In fig. \ref{fig:Va_RB}, the behavior for $^{87}Rb$ is shown; in this case $g_{3D}>0$ and the $\beta>0$ case have been maintained although the logarithmic term already offers a negative sign, so this term is responsible for inverting the value of the velocity, that is, as it expands it performs a contrast with the other two positive terms, that is these terms decay slowly. In the case of $\beta <0$, the BEC speed grows faster since there are no terms under the square root that allow the speed to decrease. Therefore, the expansion in the BEC is to infinity. That means  that there is a possibility of finding some quantum droplets in the region of $\beta>0$.

\begin{figure}[htbp!]
\centering
   \includegraphics[width=0.8\linewidth]{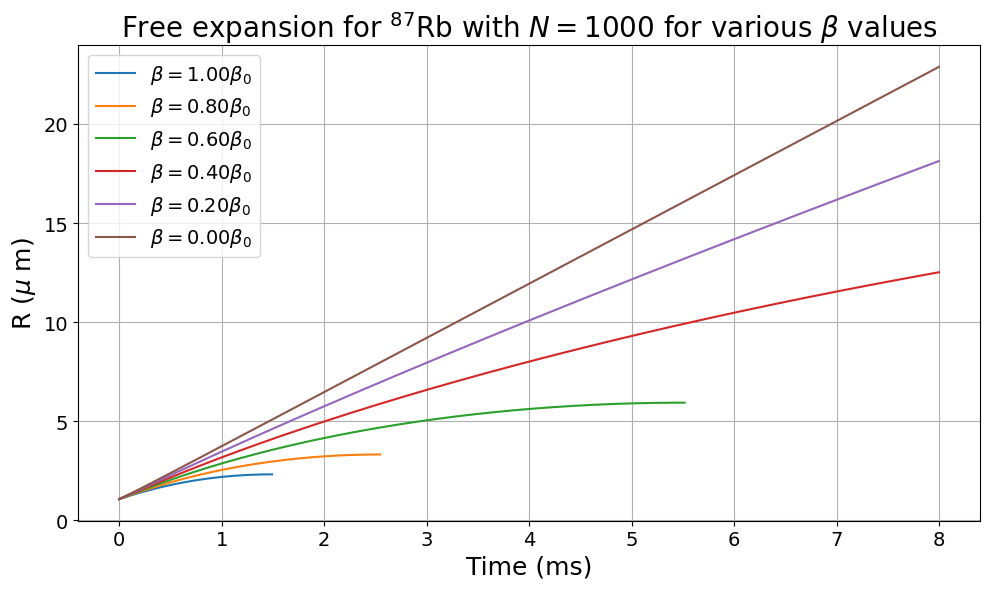}
    \caption{$^{87}Rb$: The expansion of $^{87}Rb$ is due to the fact that the logarithmic interaction energy is lower compared to its kinetic and interatomic counterpart, for $\beta <0$ it is impossible to have a confinement, however, in the region $[0, \beta_0]$ it is possible to find a confinement in the expansion of $^{87}Rb$. It is very similar to the case of $^{27}Na$ because $g_{3D}>0$ and $\beta>0$ as $^{87}Rb$.}
    \label{fig:R_FE_RB}
\end{figure}

\begin{figure}[htbp!]
    \centering
    \includegraphics[width=0.8\linewidth]{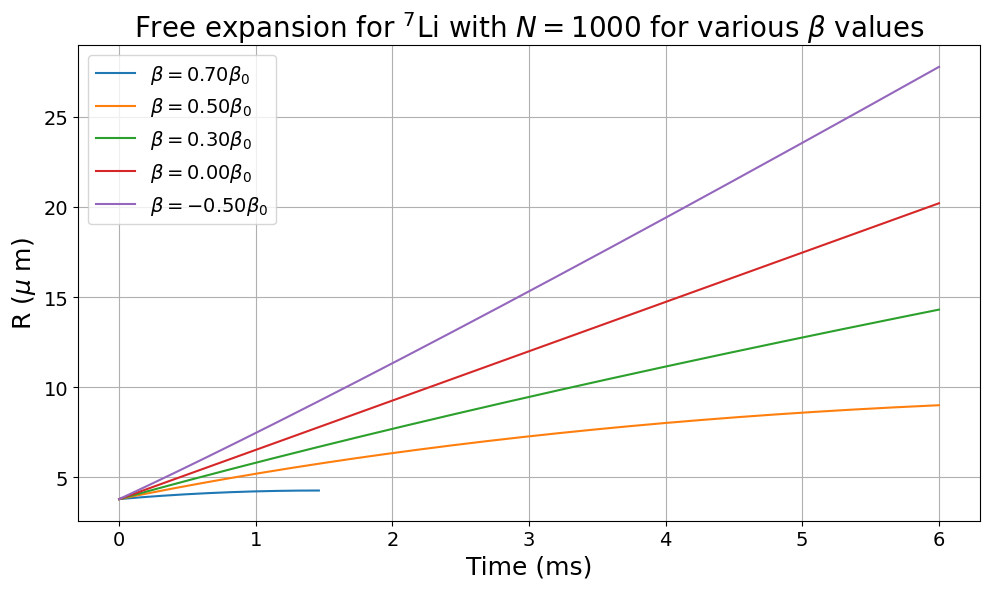}
    \caption{$^{7}Li$: The difference between the $^{87}Rb$ and $^{27}Na$ cases is that $^{7}Li$ has a negative dispersion length, which makes the interaction energy term negative and therefore attractive, so the stability region is very sensitive to $\beta$ values. For the case $\beta<0$ there is no stability and for the cases $\beta >0$, the stability is in a very short region compared to the counterpart of Rb or Na.}
    \label{fig:R_FE_Li}
\end{figure}

There are some differences between the 3D BEC results in the context of LogSE, in Ref. \cite{rodriguez2021oscillating}, the same system was analyzed, only that an anisotropic trap was used and the magnetic trap was turned off in the \textit{z} direction, verifying the existence of oscillations around an equilibrium radius. The system may be radially contained but in the \textit{z} direction it showed oscillations. In this work, the magnetic trap is turned off in all directions, checking the free expansion it is observed that under certain conditions the system can "self-contain" towards a new equilibrium radius as shown in the Fig. \ref{fig:R_FE_RB} and Fig. \ref{fig:R_FE_Li}

Another work that carried out a theoretical study and experimental contrast is found in Ref. \cite{ernst1998free}, in such research, areas are found where the expansion of the BEC approaches a constant rate, that is, a radius that limits the size of the BEC. It is not so different from those calculated here in the context of LogSE \cite{gorlitz2001realization, RevModPhys.81.647, PhysRevLett.116.215301, zloshchastiev2012volume, rodriguez2021oscillating}. 
\subsection{Rotating BEC in 3D}
\noindent When the BEC starts to rotate, it is important to choose an Ansatz that maintains a certain convergence in the region of interest. The centrifugal force that appears has a term $r^{-2}$ in the energy functional. Therefore, we propose an ansatz of the ground state of a 3D oscillatory quantum system, defined as in section \ref{sec: BEC3D-NotRot}, and the energy functional for the 3D and rotating case is defined by using the following statements:

Suppose a system of a Bose-Einstein condensate, in a magnetic trap and reaching the BEC state, a rotation is applied to the problem with an angular velocity $\Omega$ and in the radial direction, which implies that it only has angular momentum in the \textit{z} direction, that is $-\Omega L_z$, and Eq. \ref{Eq:QD} becomes

\begin{eqnarray}
&&i\hbar \frac{\partial \Psi(\mathbf{r}, t)}{\partial t} = \nonumber \\
&&\left[ -\frac{\hbar^2}{2m} \nabla^2 + V_{\text{trap}}(\mathbf{r}, t) + g_{3D}|\Psi(\mathbf{r}, t)|^2 - \beta \ln\left(\alpha^3 |\Psi(\mathbf{r}, t)|^2\right) - \Omega L_z \right] \Psi(\mathbf{r}, t),
\label{eq:QDROT-BEC-3D}
\end{eqnarray}
with $L_z = -i\hbar \frac{\partial}{\partial \theta}=i \hbar \qty(y\frac{\partial}{\partial x}-x\frac{\partial}{\partial y})$.

Then, the energy functional is given by:

\begin{equation}
E[\Psi] = \int d\mathbf{r}^3 \left( \frac{\hbar^2}{2m} |\nabla \Psi|^2 + V_{\text{trap}} |\Psi|^2 + \frac{g_{3D}}{2} |\Psi|^4 - \beta |\Psi|^2\ln\left(\alpha^3 |\Psi|^2\right) - \Omega \Psi^\ast L_z \Psi \right).
\end{equation}

In this scenario, the problem of the existence of quantum droplets in 3D will be analyzed. The rotation around the \textit{z}-axis will be used; this means that we do not work in spherical coordinates but in cylindrical coordinates instead. First, the mathematical analysis of the corresponding equation of motion will be done, the cylindrical shape like a cigar ($\omega_z << \omega_\perp$) will be analyzed and then the shape like a pancake ($\omega_z >> \omega_\perp$), the symmetrical case will also be analyzed ($\omega_z = \omega_\perp$)

We will keep the same BEC parameter, but this time it is multiplied by a complex function that determines the vorticity of the system. 

Thus:
\begin{equation}
    \Psi(r,z,\theta, t) = A re^{\frac{-r^2}{2a^2}}e^{\frac{-z^2}{2b^2}}e^{iq\theta}e^{-i\omega_rt}.
    \label{Eq: ansatz}
\end{equation}

Naturally, in Eq.~\ref{Eq: ansatz}, we set \( a = \sqrt{\frac{\hbar}{m\omega_r}} \) and \( b = \sqrt{\frac{\hbar}{m\omega_z}} \). As a result, certain differences arise in the energy functional, which is also subject to the normalization condition \( \int |\Psi|^2 \, d^3r = N \). In this case, we find \( A = \sqrt{\frac{N}{\pi^{3/2} a^4 b}} \). Additionally, the radial component of the ansatz must include at least a linear dependence on \( r \), because—although the angular part of the kinetic energy is convergent—the radial contribution diverges in the absence of such a term.

 Consequently, the log term is reduced as
\begin{equation}
    -\beta\ln (\alpha^3\abs{\psi(r)}^2)=-\beta \qty[\ln(\frac{\alpha^3 Nr^2}{\pi^{3/2} a^4 b})-\frac{r^2}{a^2}-\frac{z^2}{b^2}].
\end{equation}
    
\noindent Once again, for practical purposes, the methodology applied in the non-rotating case can be reduced by energy conservation using a posterior ansatz of the form: $\Phi=\Psi'(r,z,\theta, t)e^{i\eta(r,z)}$. 

Thus, we obtain for the total energy:
\begin{equation}
    E\qty[\Phi]-E\qty[\Psi]=E_{flow}+E_{kin}'-E_{kin}+E_{int}'-E_{int}+E_{log}'-E_{log}=0.
\end{equation}

Leading to:
\begin{eqnarray}
    E_{flow}=E_{kin}-E_{kin}'&+&E_{int}-E_{int}'+E_{log}-E_{log}' =E_{initial}-E_{final}\nonumber \\
    \frac{\hbar^2}{2m}\int \abs{\Psi'}^2 \abs{\grad \eta}^2 d\mathbf{r}^3 &=&\Delta \qty[\int d\mathbf{r}^3 \left( \frac{\hbar^2}{2m}|\nabla \Psi|^2 + \frac{g_{3D}}{2}|\Psi|^4 - \beta\qty(|\Psi|^2\ln\left(\alpha^3 |\Psi|^2\right)) - \Omega \hbar q \abs{\Psi}^2 \right) ].\nonumber
\end{eqnarray}
The delta symbol $\Delta$ establishes the difference between the initial state minus the final state. The final state can be manipulated; that is, certain physical properties can be kept constant to find more stable solutions.

Then, the corresponding energies associated with the system are given by:

\begin{eqnarray}
    E_{kin}&=&\int d\mathbf{r}^3 \qty(\frac{\hbar^2}{2m}|\nabla \Psi|^2)=\frac{\hbar^2N}{2ma^2}+\frac{\hbar^2q^2N}{2ma^2}+\frac{\hbar^2N}{4mb^2} \nonumber \\
    E_{trap}&=&\int d\mathbf{r}^3 \qty(\frac{1}{2}m\omega_rr^2+\frac{1}{2}m\omega_zz^2)\abs{\Psi^2}=mN \qty(\omega_ra^2+\frac{\omega_z}{4}b^2)
    \nonumber \\
    E_{int}&=&\int d\mathbf{r}^3 \qty(\frac{g_{3D}}{2}| \Psi|^2)=\frac{g_{3D}N^2}{\pi^{3/2}a^2b} \nonumber \\
    E_{log}&=&-\int d\mathbf{r}^3\beta|\Psi|^2\ln\left(\alpha^3 |\Psi|^2\right)=-\beta N \qty[ln\qty(\frac{\alpha^3 N}{\pi^{3/2}a^2b})-\gamma-\frac{3}{2}] \nonumber \\
    E_{rot}&=&-\int d\mathbf{r}^3\Omega \hbar q \abs{\Psi}^2=-\Omega \hbar q N
\end{eqnarray}
where $\gamma$ is the Euler-Mascheroni constant ($\gamma \approx 0.57722$). Finally, the total energy becomes:
\begin{eqnarray}
    E&=&\frac{\hbar^2N}{2ma^2}+\frac{\hbar^2q^2N}{2ma^2}+\frac{\hbar^2N}{4mb^2}\nonumber \\
    &+&mN \qty(\omega_ra^2+\frac{\omega_z}{4}b^2)+\frac{g_{3D}N^2}{\pi^{3/2}a^2b}-\beta N \qty[ln\qty(\frac{\alpha^3 N}{\pi^{3/2}a^2b})-\gamma-\frac{3}{2}]-\Omega \hbar q N.
\end{eqnarray}

The value of $q$ that minimizes the energy of the system is related to $q=\Omega/\omega_r$, indicating that when the oscillation frequency of the trap coincides with the rotation frequency, then the energy reaches its minimum. In order to calculate the free expansion, we separate the system into two parts: the disk shape and the cigar shape.

\subsection{Free expansion in a disk-shaped configuration}

A disk--shaped BEC arises when the trapping potential used to confine the particles is anisotropic, created by a combination of optical or magnetic traps. The geometry is such that the trap is much tighter in one direction (\textit{z}-axis) compared to the other two (the \textit{x}- and \textit{y}-axes). In a disk-shaped BEC, the motion of the particles along the tightly confined direction is effectively frozen, reducing the dimensionality of the system to two. This quasi-2D regime is often described by an effective 2D Gross-Pitaevskii equation, where the interaction strength is modified to account for the reduced dimensionality. The condition for obtaining a disk-shaped BEC is that $\omega_r<< \omega_z$. The radial expansion velocity has the form $v=r\frac{\dot{a}}{a}$; therefore, when introduced into the flow equation, we obtain:
\begin{equation}
    E_{flow}=\frac{\hbar^2}{2m}\int \abs{\psi}^2 \abs{\grad \eta}^2 dr^3=m N \dot{a}^2.
\end{equation}
In this case, expansion is required while the height is kept constant, and the radial trapping potential is turned off. Furthermore, the system continues to rotate with the same angular velocity as initially, implying that the rotational energy remains unchanged and does not influence the subsequent dynamics.

The equation of motion for $a$ is given by:
\begin{eqnarray}
    \dot{a}^2=\frac{\hbar^2}{2m^2}\qty(\frac{1}{a_0^2}-\frac{1}{a^2})+\frac{\hbar^2q^2}{2m^2}\qty(\frac{1}{a_0^2}-\frac{1}{a^2})+\frac{g_{3D}N}{\pi^{3/2}mb_0}\qty(\frac{1}{a_0^2}-\frac{1}{a^2})-\frac{2\beta }{m} \ln\qty(\frac{a}{a_0}).\,\,
    \label{Eq:dota_QD-3D}
\end{eqnarray}

Eq. \ref{Eq:dota_QD-3D} remains valid as long as the height of the BEC remains constant and its angular velocity is maintained at all times. However, one may also consider a scenario in which the rotation of the BEC is activated only after the trapping potential is turned off. In that case, the equation must be modified accordingly, leading to the following:

\begin{eqnarray}
    \dot{a}^2&=&
    \frac{\hbar^2}{2m^2}\qty(\frac{1}{a_0^2}-\frac{1}{a^2})+\frac{\hbar^2q^2}{2m^2}\qty(\frac{1}{a_0^2}-\frac{1}{a^2}) +\frac{g_{3D}N}{\pi^{3/2}mb_0}\qty(\frac{1}{a_0^2}-\frac{1}{a^2})\nonumber \\
    &-&\frac{2\beta }{m} \ln\qty(\frac{a}{a_0})-\frac{\Delta\Omega \hbar q}{m},
    \label{Eq:dota-QD-3D-ROT}
\end{eqnarray}
where $\Delta \Omega=\Omega_i-\Omega_f$. Both scenarios can be considered to explore interesting properties of the system. Depending on the sign of \( \Delta \Omega \), the extent of the free expansion can either increase or decrease, depending on whether the system starts with a given angular velocity and ends without it, or vice versa—starting without rotation and acquiring angular velocity later.

\begin{figure}[htbp!]
    \centering
        \centering
        \includegraphics[width=\linewidth]{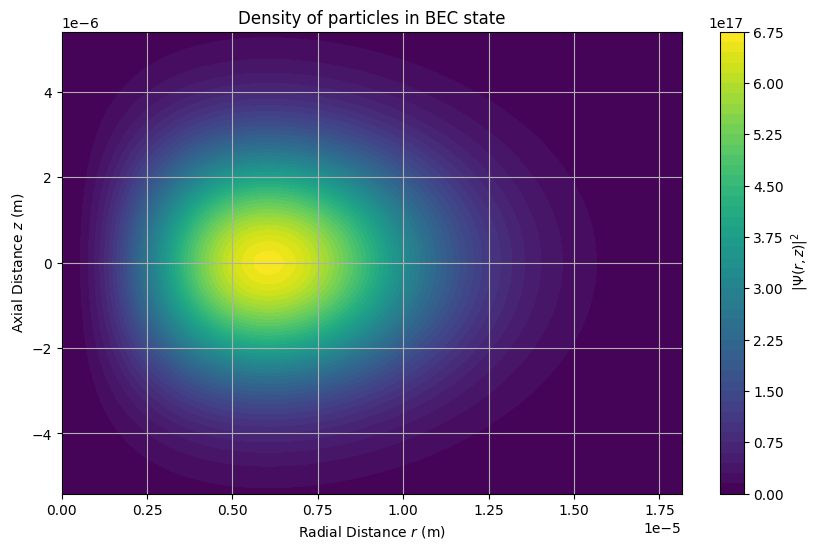}
        \caption{}
        \label{fig:subfig1}
\end{figure}

\begin{figure}[htbp!]
    \centering
        \centering
        \includegraphics[width=\linewidth]{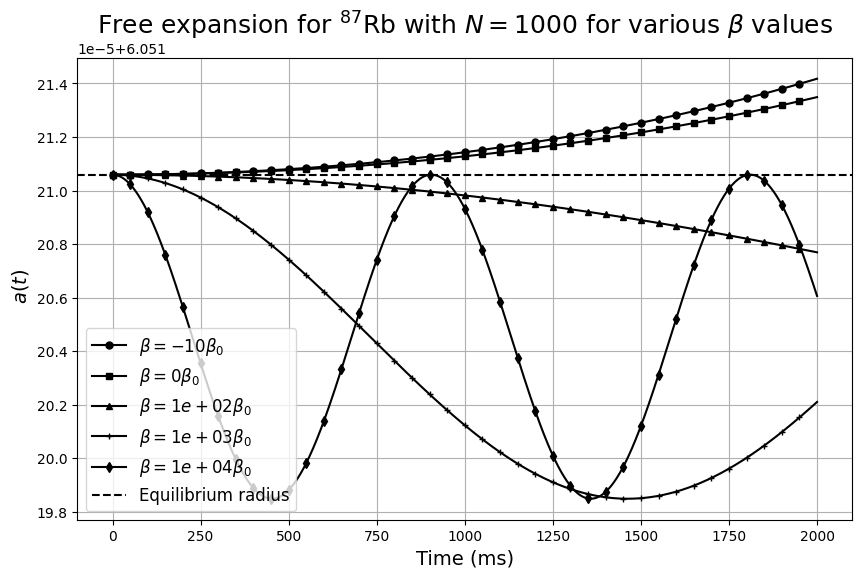}
        \caption{The solution of Eq. \ref{Eq:dota_QD-3D} is presented for different values of $\beta$. We use the RK4 method and plot the solution for the time interval from 0 to 2 seconds. The behavior of $^{27}Na$ is similar, and the simulation was performed for a disk-shaped configuration. The dashed line corresponds to the case without turning off the magnetic-optical trap. For values greater than $100\beta_0$, there is the possibility of self-confinement, leading to the formation of quantum droplets as Ref. \cite{zloshchastiev2012volume, rodriguez2021oscillating, vowe2020detecting} }
        \label{fig:Free-Expansion-RB-beta}
\end{figure}

For $N=1000$ $^{87}Rb$ atoms, Eq. \ref{Eq:dota-3DQD} has been solved at different final times depending on the plot presented in Fig. \ref{fig:subfig1}-\ref{fig:Free-Expansion-RB-beta}. The perpendicular frequency must be higher than the radial frequency to obtain a disk-shaped BEC. Then, the values $\omega=10\, rad/s, \,\omega_{z}=10\,\omega$ and $\omega_{\perp}=2\omega$ show that the height is $2.05 \,\mu m$ and the radius of the disk is $6.05 \mu m$. Although the system is not perfectly flattened, our focus is solely on the radial evolution. The radial dynamics of the BEC disk have been studied for \({}^{87}\text{Rb}\), with a time step of \(dt = 0.001\) used for temporal evolution. The radial component was adjusted in micrometers using the RK4 method.
It can be observed that for different times in milliseconds, each scheme evolves differently depending on whether there is a $\beta$ interaction or an angular velocity interaction. In addition, we use $\beta_0=\hbar \omega_\perp$ to compare different scenarios and $\Delta\Omega=1000 rad/s$.

In Fig. \ref{fig:subfig1} the density of particles in the BEC state for $^{87}Rb$ atoms is shown; it can be noted that in the center it has a very high density. 

In Fig. \ref{fig:Free-Expansion-RB-beta} from the case $\beta<0$, it can be observed that the radius of the disk tends to grow without limit; this same behavior is also observed with the case $\beta=0$. It can also be observed that for values of $\beta >0$ there may be oscillations below the equilibrium radius, that is, through the existence of quantum droplets for $^{87}Rb$. 

\begin{figure}[htbp!]
    \centering
        \centering
        \includegraphics[width=\linewidth]{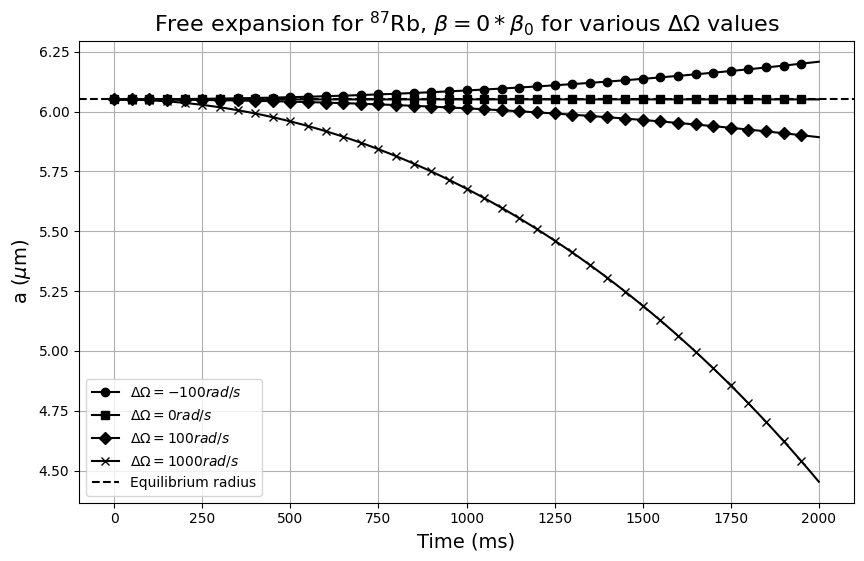}
        \caption{Rotation of the $^{87}Rb$ atoms in BEC state is presented. We can see that the $\beta=0$ in order to see a rotation of a BEC without LogSE model. Quantum droplets are not observed, in which the Bose-Einstein condensate remains localized around some equilibrium radius.}
        \label{fig:Rot-beta0}
\end{figure}

\begin{figure}[htbp!]
    \centering
        \centering
        \includegraphics[width=\linewidth]{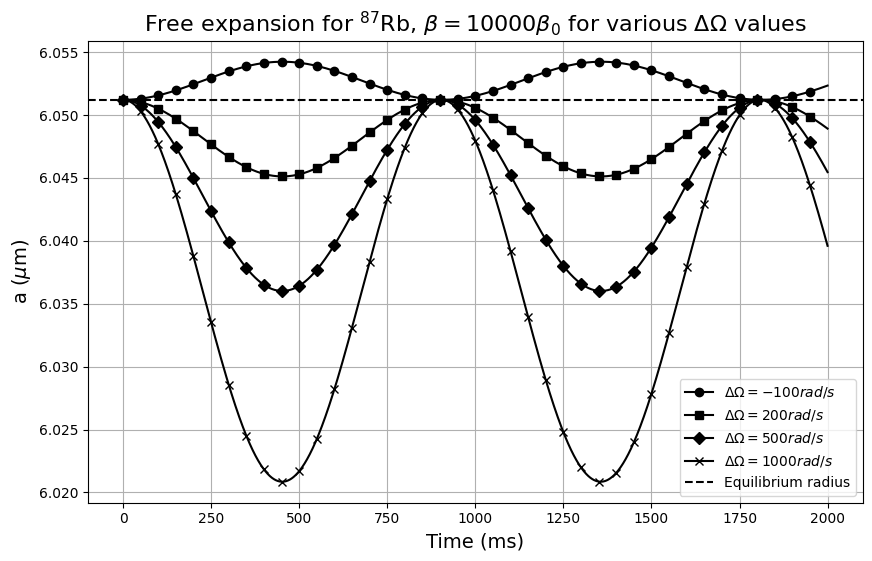}
        \caption{The rotation of the $^{87}\text{Rb}$ atoms in the BEC state is presented. We can see that the strength of $\beta = 10000\beta_0$ implies that the $\beta$ values prevent collapse or infinite expansion, and in rotation, it allows oscillations in specific regions.Quantum droplets are  observed, in which the Bose-Einstein condensate remains localized around some equilibrium radius.}
        \label{fig:Rot-beta1000}
\end{figure}

In Fig. \ref{fig:Rot-beta0} we can observe that when $\Delta \Omega <0$, there is an excessive increase in the speed and therefore in the expansion of the radius of the disk. However, it continues oscillating for the corresponding $\beta$ value. In the opposite case in Fig. \ref{fig:Rot-beta1000}, the radius of the disk is compressed for the corresponding  $\beta$ value.

%%%%%%%%%%%%%%%%%%%%%%%%%%%%%%%%%%%%%%%%%%%%%%%%%%%%%%%%%%%%%%%%

 In the case of $^{27}Na$, the oscillatory behavior is similar. However, for $^{7}Li$, we observe that it is highly sensitive to variations in \(\beta\) and \(\omega\). Furthermore, for $^{87}Rb$, if \(\Delta \Omega \ll -1 \times 10^5 \, \text{rad/s}\), an increasing velocity is observed, allowing the BEC to expand freely with \(\beta < 100\beta_0\). In this case, we may obtain a cloud of $^{87}Rb$, and no further BEC states are present in that configuration. It is important to note that \(\Delta \Omega\) can be negative because the initial angular velocity may be much smaller than the final angular velocity.The angular component of the energy of the quantum droplets can exhibit two behaviors: free expansion or compression to a new equilibrium radius, as shown in Fig. \ref{fig:Rot-beta0} and Eq. \ref{Eq:dota-QD-3D-ROT}.

\section{Summary and conclusions}
The existence of quantum droplets is observed in positive $\beta$ regions from Fig. \ref{fig:subfig1}-\ref{fig:Rot-beta1000}. The comparison of an expanding sphere is only expandable up to a certain limit. However, quantum droplets can be found from self-confined BECs. From  Fig. \ref{fig:R_FE_RB}, we can observe that the system looks like a in a free expansion but stops at a certain volume. It cannot be assured whether it has expanded and stopped due to $\beta$ interaction or equations only model that expansion region and then undergo a phase change to a cloud of $^{87}Rb$ atoms. However, if we fix one dimension as in the disk-shaped BEC, we can observe oscillations around an equilibrium radius, i.e., the prediction of quantum rubidium droplets. So, in Ref. \cite{rodriguez2021oscillating} did the free expansion in one dimension (\textit{z}-axis), observing oscillations in the height since it was cigar-shaped. In this work, we have used the disk-shaped in order to observe the radial expansion in $^{87}Rb$. Energy conservation is quite appropriate; since these are macroscopic properties that can be used to understand the evolution of a BEC, and its free expansion can be easily modeled. However, numerical methods used, such as RK4 method, may present a limited scheme in specific space-time regions. Let us remark that, the results obtained in the present report open up the possibility of application in many other fields. For instance, the extension of the formalism developed here can be applied in the relativistic regime. This topic deserves a deeper analysis that we will present elsewhere. Undoubtedly, quantum droplets have a vast field to explore, so research continues to be open in this area, like gravitation and cosmology, for instance, in the dark matter and perhaps dark energy schemes, see references \cite{BS1,BS2}.





\bibliography{BIB-JTP}% common bib file

\end{document}